\documentclass[a4paper]{article}
\usepackage{url}
\usepackage{INTERSPEECH2022}
\usepackage{booktabs,adjustbox,makecell,multirow}
\usepackage[colorlinks,linkcolor=red]{hyperref}
\usepackage{diagbox}

\title{The USTC-NERCSLIP Systems for the CHiME-8 NOTSOFAR-1 Challenge}
\name{Shutong Niu$^1$, Ruoyu Wang$^1$, Jun Du$^{1,*}$\thanks{* Corresponding author}, Gaobin Yang$^1$, Yanhui Tu$^2$, Siyuan Wu$^2$, Shuangqing Qian$^2$, Huaxin Wu$^2$, Haitao Xu$^2$, Xueyang Zhang$^2$, Guolong Zhong$^2$, Xindi Yu$^2$, Jieru Chen$^2$, \\ Mengzhi Wang$^2$, Di Cai$^2$, Tian Gao$^2$, Genshun Wan$^2$, Feng Ma$^2$, Jia Pan$^2$, Jianqing Gao$^2$}
\address{
  $^1$University of Science and Technology of China, China \\$^2$iFlytek Research, China}
\email{\{niust,wangruoyu\}@mail.ustc.edu.cn, jundu@ustc.edu.cn}

\begin{document}

\maketitle
\begin{abstract}
This technical report outlines our submission system for the CHiME-8 NOTSOFAR-1 Challenge \cite{vinnikov24_interspeech}. The primary difficulty of this challenge is the dataset recorded across various conference rooms, which captures real-world complexities such as high overlap rates, background noises, a variable number of speakers, and natural conversation styles. To address these issues, we optimized the system in several aspects: For front-end speech signal processing, we introduced a data-driven joint training method for diarization and separation (JDS) to enhance audio quality. Additionally, we also integrated traditional guided source separation (GSS) for multi-channel track to provide complementary information for the JDS. For back-end speech recognition, we enhanced Whisper with WavLM, ConvNeXt, and Transformer innovations, applying multi-task training and Noise KLD augmentation, to significantly advance ASR robustness and accuracy.
Our system attained a Time-Constrained minimum Permutation Word Error Rate (tcpWER) of 14.265\% and 22.989\% on the CHiME-8 NOTSOFAR-1 Dev-set-2 multi-channel and single-channel tracks, respectively.

\end{abstract}
\noindent\textbf{Index Terms}: CHiME challenge, speaker diarization, speech separation, speech recognition, joint training

%\section{Introduction}

\section{System Description}
\begin{figure}[htb]
	\centering
	\includegraphics[width=\linewidth]{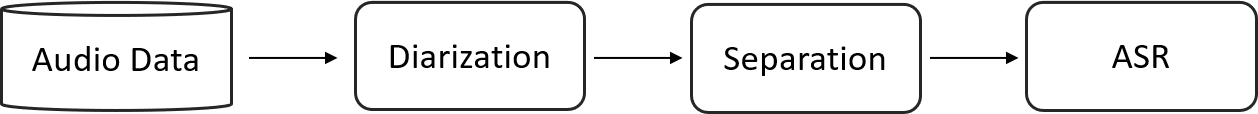}
	\caption{ Overall framework of the system.}
	\label{fig:1}
\end{figure}

Our overall system follows the process illustrated in Fig. \ref{fig:1}. First, the diarization system is used to predict the speaker's time distribution, which is then utilized to perform speech separation. Then, the separated speech is sent to the speech recognition system. In the following sections, we will describe the single-channel and multi-channel systems in detail.
\subsection{Multi-channel System}
\subsubsection{Diarization}
\begin{figure}[htb]
	\centering
	\includegraphics[width=0.5\textwidth]{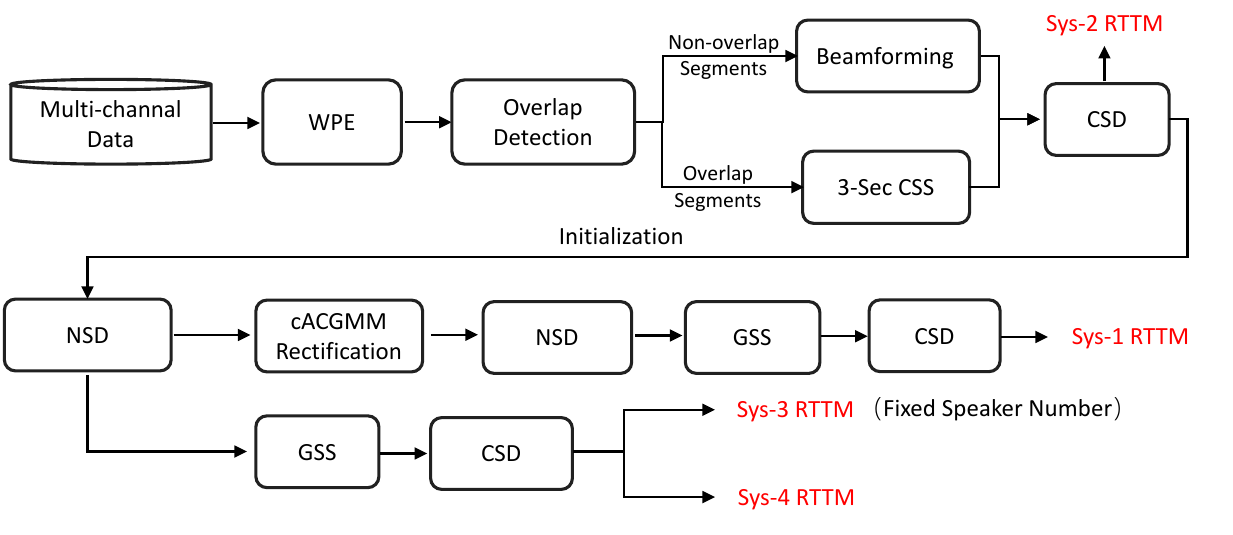}
	\caption{The diarization pipeline for multi-channel system.}
	\label{fig:2}
\end{figure}
Fig. \ref{fig:2} illustrates the diarization component of the multi-channel system. For the original multi-channel data, we first perform weighted prediction error (WPE) algorithm, followed by overlap segment detection. We used the same architecture as the separation model of the official CSS baseline \cite{chime8basline} for the overlapping segment detection model. However, we changed the sliding window length to 800 frames (12.8 seconds) and modified the final prediction output to the frame-level binary classification using a linear layer. For the detected overlapping segments, we employ the multi-channel 3-second continuous speech separation (CSS) method to effectively isolate each speaker's speech. We modified the official baseline architecture \cite{chime8basline} for the CSS on overlapping segments by adding a classification network for overlapping segment detection and conducting joint training for separation and overlap segment detection. We used the official model for initialization and conducted joint training to enhance the separation model's ability to differentiate between overlapping and non-overlapping segments. We only used the predicted results from the separated parts. The sliding window length was kept at 3 seconds. We used the same training and inference procedures as the official baseline. The training data remained consistent with the baseline, utilizing only the official simulated data \cite{chime8simulate}. For non-overlapping segments, we enhance the multi-channel speech using the MVDR beamformer \cite{higuchi2017online}. 

We conduct the clustering-based speaker diarization (CSD) method on these pre-processed speech, resulting in preliminary speaker diarization priors, referred to as `Sys-2 RTTM' in Fig. \ref{fig:2}. For CSD system, we use the spectral clustering algorithm. We leverage the ResNet-221 model for speaker embedding extraction, which is trained on the VoxCeleb \cite{nagrani2020voxceleb} and LibriSpeech datasets. To obtain different diarization priors, we further apply various processing techniques to the speech used for clustering. Firstly, we use the results obtained from clustering as initial priors, and feeding them into the neural network-based speaker diarization (NSD) system to achieve more precise speaker boundary information. The NSD employed in our system is the memory-aware multi-speaker embedding with sequence-to-sequence architecture (NSD-MS2S) \cite{MAMSE, yang2024neural}, which combines the advantages of memory-aware multi-speaker embedding and sequence-to-sequence architecture. For the multi-channel track, we input different channels separately and then averaged the posterior probabilities of the different channels to obtain the final result for one session. The NSD uses the 800-frame window length with a frame length of 10ms, resulting in a total window length of 8 seconds. Then, following our previous methods in CHiME-7 DASR Challenge \cite{Wang2023TheUS}, we conduct cACGMM rectification on the original audios, adopting a window length of 120 seconds and a window shift of 60 seconds. This rectification utilizes the previous NSD decoding result as the initialization mask. By implementing a threshold on the spectrum mask of the cACGMM, we obtain a refined secondary initialization of diarization results for the NSD system. After the official GSS initialized with the second NSD decoding results, we perform the re-clustering to obtain better diarization priors (Sys-1 RTTM). Additionally, the decoding results from the first NSD can be directly utilized to initialize the GSS, thereby generating separated audios. For these separated audios, we conducted re-clustering with the fixed number of speakers (maintaining the global number of speakers within a session) and the non-fix number of speakers (the original version), resulting in two initial diarization priors, namely `Sys-3 RTTM' and `Sys-4 RTTM', respectively.

\subsubsection{Separation}
\begin{figure}[tb]
	\centering
	\includegraphics[width=0.45\textwidth]{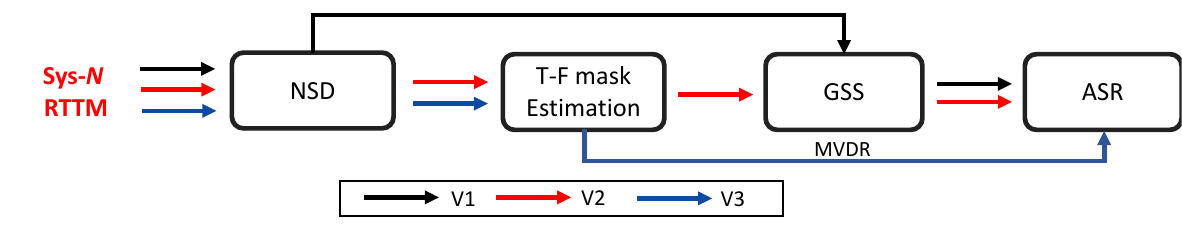}
	\caption{The separation pipeline for multi-channel system.}
	\label{fig:3}
	\vspace{-0.5 cm}
\end{figure}

\begin{figure}[tb]
	\centering
	\includegraphics[width=0.95\linewidth]{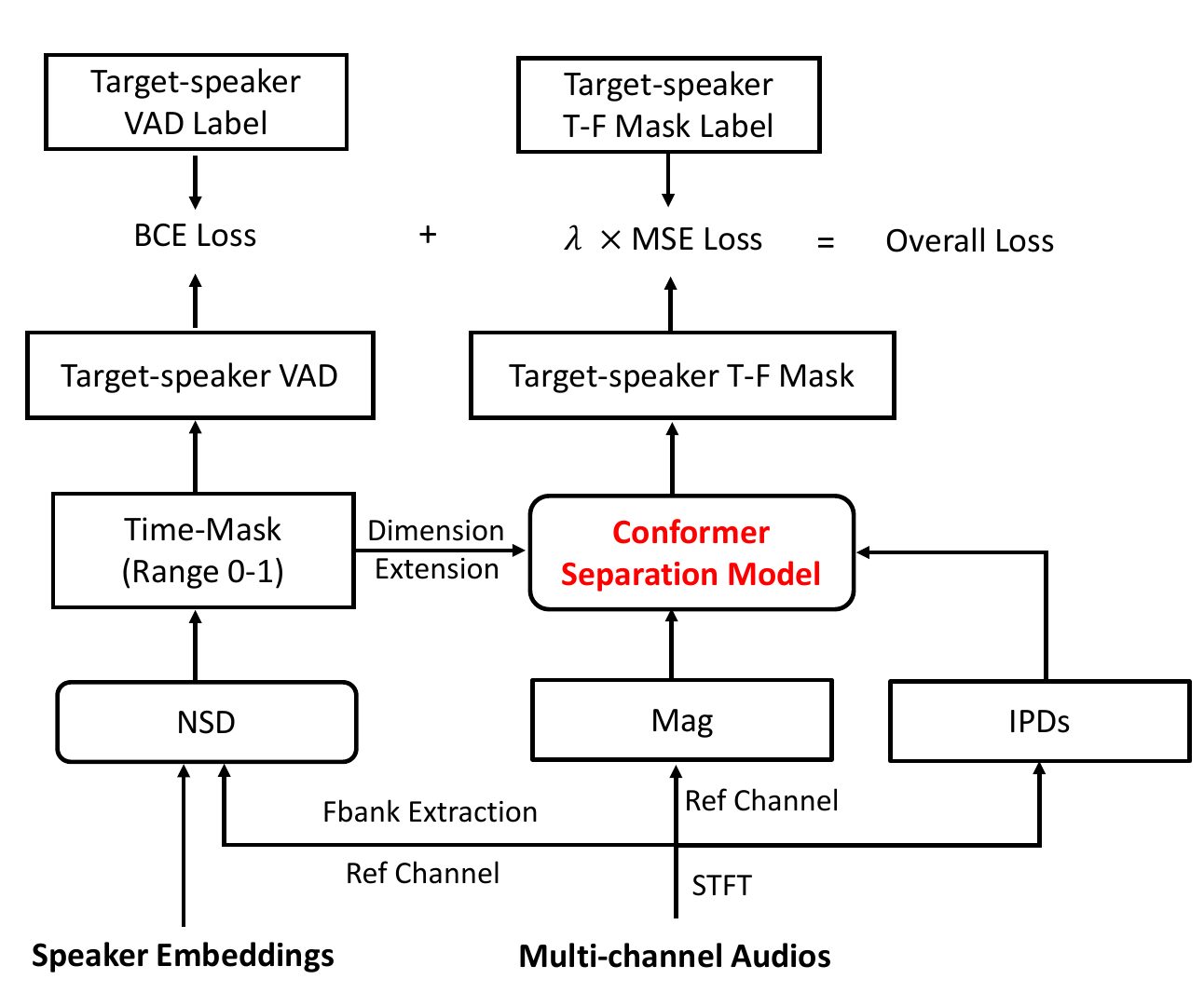}
	\caption{Overall framework of multi-channel JDS method.}
	\label{fig:4}
\end{figure}

After obtaining the RTTMs from the diarization system, we acquire information about the speaker distribution. Utilizing this information, we proceed with various versions of speech separation as depicted in Fig. \ref{fig:3}. For the first system (V1), we utilize the NSD to optimize the time boundaries. The optimized results are then used to initialize the GSS algorithm, resulting in the separated audios. For the second system (V2), we utilize the time masks estimated from the NSD as the inputs for JDS system. This guides the JDS system in estimating time-frequency (T-F) soft masks. These T-F masks are then employed to initialize the GSS in the T-F domain, thus providing the GSS with initialization information in both time and frequency dimensions. For the third system (V3), we directly utilize the T-F masks predicted by the JDS system to guide the MVDR beamforming, while still employing the time boundaries provided by the NSD to get the separated speech segments. Fig. \ref{fig:4} shows the overall framework of multi-channel joint training method for diarization and separation (JDS). The JDS system comprises two main components: the speaker diarization module and the speech separation module. Based on the original end-to-end speaker diarization systems, the JDS system serially integrates the separation module. This helps the speech separation system accurately identify the number of speakers and the corresponding identities. This information also facilitates the speech separation model in more effectively distinguishing between different speakers. Consequently, the separation module in JDS system primarily maps the time information of various speakers to time-frequency information, which significantly simplifies the speech separation process. In our system, the JDS uses a window length of 800 frames with a frame length of 16ms, resulting in a total window length of 12.8 seconds.
% \begin{table}[htb]
	% \caption{Different versions of speech separation for multi-channel system.}
	% \label{tab:1}
	% \centering
	% \begin{tabular}{c|c}
		% \toprule
		% \textbf{Version Index} & \textbf{Separation Method} \\
		% \midrule
		% V1 & NSD + GSS \\
		% V2 & NSD + T-F mask Estimation (JDS) + GSS \\
		% V3 & NSD + T-F mask Estimation (JDS) + MVDR \\
		% V4 & GSS  \\
		% V5 & T-F mask Estimation (JDS) + GSS \\
		% V6 & T-F mask Estimation (JDS) + MVDR \\
		% \bottomrule
	% \end{tabular}
% \end{table}

\subsubsection{Speech Recognition}

\begin{figure}[t]
	\centering
	\includegraphics[width=0.7\linewidth]{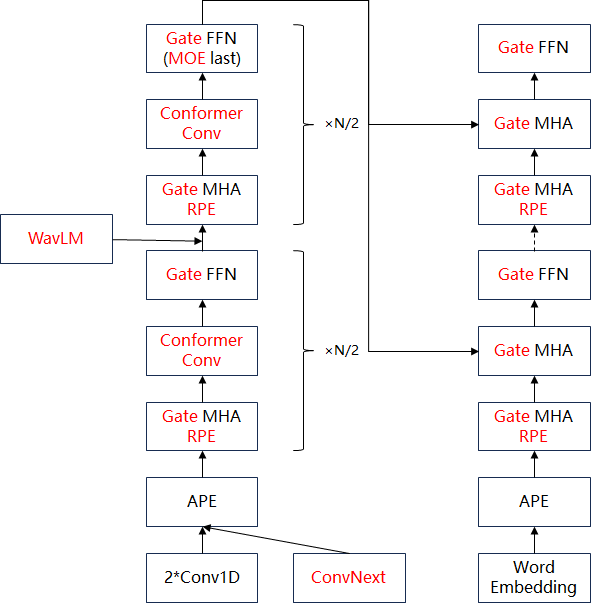}
	\caption{The architecture of Enhanced Whisper.}
	\label{fig:Enhanced Whisper}
\end{figure}

For automatic speech recognition tasks, we leverage Whisper \cite{Radford2022RobustSR}, a state-of-the-art open-source model renowned for its high accuracy. Whisper follows an encoder-decoder architecture based on the Transformer framework. The input to the model is represented as a log Mel-spectrogram. Both the encoder and decoder components feature absolute positional encoding and are composed of several transformer layers. Notably, the encoder contains two layers of 1D convolution preceding the absolute positional encoding stage, which aids in extracting local features from the input audio data.

We introduce Enhanced Whisper, a variant that introduces a series of enhancements to the base Whisper model. An overview of the modified architecture is illustrated in Fig. ~\ref{fig:Enhanced Whisper}. To refine input feature representation, we drew inspiration from the CHiME-7 DASR Challenge \cite{Wang2023TheUS}, leveraging features extracted from self-supervised pre-trained models, particularly WavLM \cite{Chen2021WavLMLS}. Our experiments involved systematically integrating these WavLM-derived features at various stages within the Whisper encoder, including the initial, intermediate, and final layers. We observed that injecting these features at the intermediate layer of the encoder resulted in a slight yet noticeable improvement in performance. The outputs from WavLM and the intermediate layer of the Whisper encoder are integrated via a concatenation operation, followed by a linear transformation to ensure compatibility with the original feature dimensions of the model.
Concerning downsampling convolutions, the baseline Whisper model utilizes two layers of 1D convolution. Inspired by recent advancements like NextFormer \cite{Jiang2022NextformerAC}, we augmented the original Whisper model with a ConvNeXt structure, running in parallel to the standard 1D convolutions. The ConvNeXt output is added to the original Conv1D output after a linear transformation and then input into the transformer.

Regarding positional encoding, Whisper initially relies on absolute positional encoding. However, empirical evidence suggests that absolute positional encoding exhibits limitations in robustness compared to relative positional encoding \cite{Zhou2019ImprovingGO}. Motivated by these findings, we adopted bias relative positional encoding \cite{Dai2021CoAtNetMC} within our enhanced model, aiming to improve its resilience and performance consistency across varying input lengths.

In terms of the Transformer block, we took cues from relevant research \cite{Sun2023RetentiveNA, Peng2023RWKVRR} to integrate a sigmoid gating mechanism. Specifically, the input is projected through a weight matrix (W), followed by a sigmoid activation function. The output of this operation is then scaled by a factor of 2 before being element-wise multiplied with the original output, effectively controlling the flow of information within the Transformer block.
Additionally, we explored the insertion of a depthwise convolution module, akin to those featured in Conformer \cite{Gulati2020ConformerCT} models, following the Multi-Head Attention (MHA) layers. This architecture enhances the model’s ability for localized modeling.
Furthermore, we augmented the final layer of the encoder with a Mixture of Experts (MoE) \cite{You2021SpeechMoEST} component, aimed at enhancing the model’s representational capacity.

\subsection{Single-channel System}
\begin{figure}[t]
	\centering
	\includegraphics[width=1.0\linewidth]{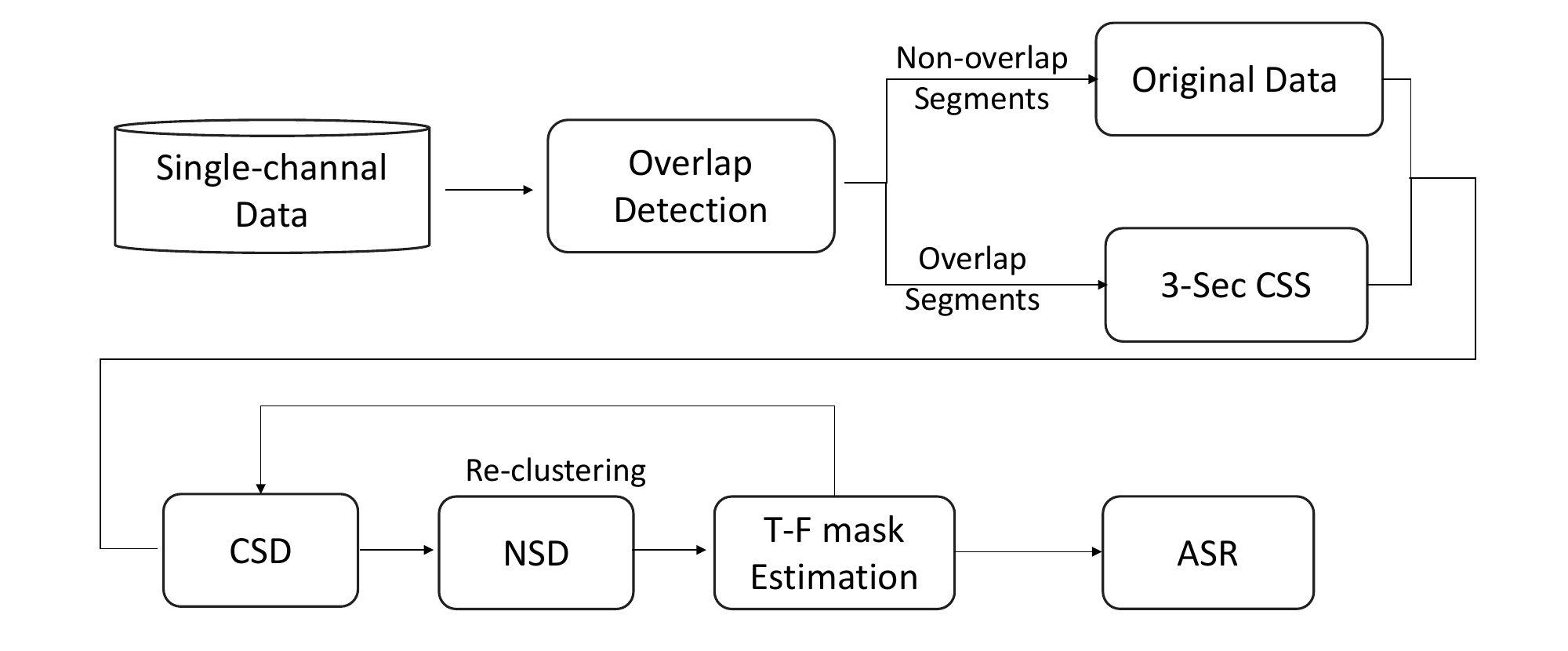}
	\caption{The framework of single-channel system.}
	\label{fig:6}
\end{figure}
The framework of the single-channel system is illustrated in Fig. \ref{fig:6}. Like the multi-channel system, the single-channel system also begins with speaker diarization followed by speech separation and ASR. However, unlike the multi-channel system, each module in the single-channel system (including the overlap detection, CSS, CSD, NSD, and T-F mask estimation) receives only single-channel audios or features as inputs. To get the separated audios for each speaker, the amplitude spectral features of the original mixed audio are multiplied by the T-F masks, and an inverse STFT transformation is performed. Furthermore, re-clustering the separated audios can enhance the precision of the speaker diarization priors, as illustrated at the bottom of the Fig. \ref{fig:6}. For ASR, we use the same model for decoding as in the multi-channel system.

\subsection{Datasets}
\label{Data_Augmentation}
\subsubsection{Diarization and Separation}
For the speaker diarization system, the training data comprises the officially simulated training dataset \cite{chime8simulate}, Train-set-1 \cite{chime8Meetingsrecordingsdataset}, Train-set-2 \cite{chime8Meetingsrecordingsdataset} and Dev-set-1 \cite{chime8Meetingsrecordingsdataset}. We also employed LibriSpeech, MUSAN noise \cite{musan2015} and the noises in officially simulated training dataset \cite{chime8Meetingsrecordingsdataset} to simulate the diarization training data\footnote{https://github.com/jsalt2020-asrdiar/jsalt2020\_simulate}. Additionally, we also use the near-field recordings from Train-set-1, Train-set-2 and Dev-set-1 as clean data to simulate multi-channel speaker diarization training data. For speech separation, we use the officially simulated training dataset and also use the near-field recordings from Train-set-1, Train-set-2 and Dev-set-1 to simulate the separation training data.

\begin{table}[t]
	\caption{The training sets of speech recognition.}
	\label{tab:data}
	\centering
	\resizebox{8 cm}{!}{\begin{tabular}{ c|c|c }
			\toprule
			\textbf{Duration (h)} & \textbf{Corpus} & \textbf{Sample Scale} \\
			\midrule
			14 & Train-set-1 MC GSS & 1 \\
			16 & Train-set-2 MC GSS & 1 \\
			10 & Dev-set-1 MC GSS & 1 \\
			14 & Train-set-1 MC GSS with timestamp & 1 \\
			16 & Train-set-2 MC GSS with timestamp & 1 \\
			10 & Dev-set-1 MC GSS with timestamp & 1 \\
			14 & Train-set-1 MC NN & 1 \\
			16 & Train-set-2 MC NN & 1 \\
			10 & Dev-set-1 MC NN & 1 \\
			14 & Train-set-1 MC ch0 NN & 1 \\
			16 & Train-set-2 MC ch0 NN & 1 \\
			10 & Dev-set-1 MC ch0 NN & 1 \\
			960 & LibriSpeech & 1 \\
			\bottomrule
	\end{tabular}}
\end{table}

\subsubsection{Speech Recognition}
The ASR systems were trained using official NOTSOFAR-1 training data and the open-source LibriSpeech dataset with data augmentation methods. The data augmentation methods included speed perturbation and MUSAN noise \cite{musan2015} addition. The specific composition of the training data is shown in Table~\ref{tab:data}. We utilized multi-channel (MC) data processed by both Guided Source Separation (GSS) and Neural Networks (NN), and introduced a word-level timestamp prediction task into the GSS data. Specifically, for GSS, we used oracle RTTM labels on the multi-channel data to perform GSS, resulting in separated speech segments with corresponding speaker identities and timestamps that match the ASR annotations. Through this correspondence, we matched the recognition labels to the separated results. For NN-based separation, we directly apply the JDS method for separation and segment the separated audios according to the time steps of oracle RTTM.
We found that this multitask training approach led to a slight improvement in recognition accuracy. We also adopted the practice from Whisper of providing the transcribed text from the preceding utterance as previous-text conditioning, which has noticeably improved the recognition rate. Contrary to using official single-channel (SC) data, we selected NN-processed MC channel 0 (ch0) data as our single-channel training input, observing superior performance with this choice. Drawing inspiration from the principles of RDrop \cite{Ji2023ResearchOA}, we developed a novel data augmentation technique called Noise KLD. This approach entails separately feeding both the original and augmented data samples into the model. Consistency between the model’s predictions for the original and augmented data is ensured by applying Kullback-Leibler divergence (KLD) loss as a regularizer. Through extensive experimentation, we discovered that this method outperforms conventional data augmentation strategies in terms of boosting model performance and generalization.

\section{Results}
For diarization, the training requires approximately 88 hours, and testing all sentences in Dev-set-2 takes about 1 hour. For the JDS system, training takes approximately 4 days, while testing all sentences in Dev-set-2 requires about 1 hour. For ASR, training requires about 20 hours, and testing all sentences in Dev-set-2 consumes about 6 hours. Typically, we conduct our training on A100 GPUs and perform testing on V100 or A40 GPUs.

\subsection{Overall Results}
\subsubsection{Multi-channel System}
Table \ref{tab:mc} presents the tcpWER ($\%$) of our multi-channel system on Dev-set-2, where `Sys-$N$ RTTM' corresponds to the system depicted in Fig. \ref{fig:2}, and `V$*$' corresponds to the system shown in Fig. \ref{fig:3}. For each system, we fused the posterior probabilities from three different Whisper models (enhanced large-v2, enhanced large-v3, and enhanced large-v3 trained with more data simulated from Librispeech). The enhanced Whisper models were fine-tuned using the official Whisper large v2 (enhanced large-v2) and v3 (enhanced large-v3 and enhanced large-v3 trained with more data simulated from Librispeech) parameters for initialization. The last row `Fusion' indicates the average of posterior probabilities across 9 (3 $\times$ 3) systems using the same speaker diarization priors. Finally, in the multi-channel track, we submit the fusion results of each `Sys-$N$ RTTM' (last column).
\begin{table}[h]
	\renewcommand\arraystretch{1.25}
	\newcolumntype{L}[1]{>{\raggedright\arraybackslash}p{#1}}
	\newcolumntype{C}[1]{>{\centering\arraybackslash}p{#1}}
	\newcolumntype{R}[1]{>{\raggedleft\arraybackslash}p{#1}}
	\centering
	\caption{TcpWER ($\%$) comparisons on the multi-channel track on Dev-set-2. }
	\vspace{-0.3cm}
	\label{tab:mc}\medskip
	\resizebox{8 cm}{!}{\begin{tabular}{c|c|c|c|c}
			%\hline
			\toprule[1 pt]
			%\hline
			\diagbox{Sep}{Dia}&Sys-1 RTTM&Sys-2 RTTM&Sys-3 RTTM&Sys-4 RTTM\\
			\midrule
			%\hline
			V1&14.953&14.649&15.116&14.571\\
			V2&14.911&14.595&15.086&14.547\\
			%\hline
			V3&15.577&15.160&15.703&15.018\\
			%\hline
			Fusion&14.681&14.286&14.847&14.265\\
			\bottomrule[1 pt]
	\end{tabular}}
	%\vspace{-0.3 cm}
\end{table}
\subsubsection{Single-channel System}
Table \ref{tab:sc} presents the tcpWER ($\%$) of our single-channel system on Dev-set-2. The diarization priors  are derived from NSD and re-clustering, as illustrated in Fig. \ref{fig:6}. These priors are then input into the JDS system, from which separated audio is obtained via multiplying T-F masks and amplitude spectrum. Similarly, for each subsystem, we have fused posterior probabilities from three different Whisper models (enhanced large-v2, enhanced large-v3, and enhanced large-v3 trained with more data simulated from Librispeech). 

\begin{table}[t]
	\renewcommand\arraystretch{1.25}
	\newcolumntype{L}[1]{>{\raggedright\arraybackslash}p{#1}}
	\newcolumntype{C}[1]{>{\centering\arraybackslash}p{#1}}
	\newcolumntype{R}[1]{>{\raggedleft\arraybackslash}p{#1}}
	\centering
	\caption{TcpWER ($\%$) comparisons on the single-channel track on Dev-set-2. }
	\vspace{-0.3cm}
	\label{tab:sc}\medskip
	\resizebox{5 cm}{!}{\begin{tabular}{c|c|c}
			%\hline
			\toprule[1 pt]
			%\hline
			\diagbox{Sep}{Dia}&NSD&Re-clustering\\
			\midrule
			%\hline
			JDS&24.611&22.989\\
			\bottomrule[1 pt]
	\end{tabular}}
	%\vspace{-0.3 cm}
\end{table}
\subsection{Ablation Results}
To better illustrate our system, we present some ablation experiments conducted during the challenge, along with some discussions in this section. We will focus on showing the ablation results of three main modules: diarization, speech separation, and speech recognition, respectively.

\subsubsection{Diarization}
Table~\ref{t2} presents the diarization results at different stages on Dev-set-2, where we define the stages based on the number of NSD decodings in Fig. \ref{fig:2}. The first stage corresponds to the first decoding of NSD in Fig. \ref{fig:2} and the CSD results used for initialization. The second stage refers to the second decoding of NSD in Fig. \ref{fig:2} and the cACGMM
rectification-based diarization results used for initialization. The third stage corresponds to the CSD results `Sys-1 RTTM' in Fig. \ref{fig:2}, along with the corresponding NSD results. For more detailed definitions, please refer to this paper \cite{wang2024incorporatingspatialcuesmodular}. The term `filter' refers to the process of eliminating segments that contain fewer than one word using a speech recognition model to prevent interference from incomplete or very short segments. As we can see, the introduction of CSS significantly improves the performance of the diarization in stage 1, effectively reducing the MISS errors in the CSD results. At the same time, stage 2 shows a relatively effective improvement in DER compared to stage 1, but the recognition performance actually become worsens. Finally, the filtering operation in stage 3 can effectively reduce SpkErr errors. However, the final DERs still don't show improvement compared to stage 2.
\begin{table}[t]
\centering
\caption{
Evaluation results \cite{wang2024incorporatingspatialcuesmodular} of the proposed diarization module on Dev-set-2 multi-channel track. The ASR model is based on Whisper-large-v3, fine-tuned on Train-set-1/2 datasets. Note that we removed the anomalous session `MTG-30522'.}
\setlength{\tabcolsep}{3pt} % 调整列间距
\resizebox{\columnwidth}{!}{
\begin{tabular}{l|cccc|cccc|c}
\toprule
& \multicolumn{4}{c|}{Initialization} & \multicolumn{4}{|c|}{NSD Decoding} & ASR \\
       & FA   & MISS  & SpkErr & DER   & FA   & MISS  & SpkErr & DER   & tcpWER \\
\midrule
Stage 1 (w/o CSS)           & 4.72 & 25.36 & 2.56   & 32.65 & 4.82 & 7.46  & 2.96   & 15.24 & -      \\
Stage 1                    & 5.90 & 15.17 & 2.36   & 23.43 & 4.77 & 7.37  & 2.26   & 14.40 & 12.87  \\
Stage 2                    & 7.51 & 11.67 & 1.89   & 21.07 & 4.51 & 7.00  & 2.47   & 13.97 & 14.13  \\
Stage 3 (w/o filter)        & 3.50 & 8.38  & 4.25   & 16.12 & 4.13 & 7.44  & 3.00   & 14.58 & 13.65  \\
Stage 3           & 3.71 & 13.87 & 1.50   & 19.09 & 4.50 & 7.47  & 2.21   & 14.19 & 12.83  \\
\bottomrule
\end{tabular}}
\label{t2}
\end{table}
% \begin{table}[t]
% \centering
% \caption{
% Evaluation results of the proposed diarization module on Dev-set-2. The ASR model is based on Whisper-large-v3, fine-tuned on Train-set-1/2 datasets. Note that we removed the anomalous session `MTG-30522'.
% }
% \setlength{\tabcolsep}{3pt} % 调整列间距
% \resizebox{\columnwidth}{!}{
% \begin{tabular}{l|cccc|cccc|c}
% \toprule
% & \multicolumn{4}{c|}{Initialization} & \multicolumn{4}{|c|}{NSD Decoding} & ASR \\
%        & FA   & MISS  & SpkErr & DER   & FA   & MISS  & SpkErr & DER   & tcpWER \\
% \midrule
% Oracle                     & -    & -    & -      & -     & 3.50 & 6.34  & 1.25   & 11.10 & 11.37      \\
% Stage 1 (w/o CSS)           & 4.72 & 25.36 & 2.56   & 32.65 & 4.82 & 7.46  & 2.96   & 15.24 & -      \\
% Stage 1                    & 5.90 & 15.17 & 2.36   & 23.43 & 4.77 & 7.37  & 2.26   & 14.40 & 12.87  \\
% Stage 2                    & 7.51 & 11.67 & 1.89   & 21.07 & 4.51 & 7.00  & 2.47   & 13.97 & 14.13  \\
% Stage 3 (w/o filter)        & 3.50 & 8.38  & 4.25   & 16.12 & 4.13 & 7.44  & 3.00   & 14.58 & 13.65  \\
% Stage 3           & 3.71 & 13.87 & 1.50   & 19.09 & 4.50 & 7.47  & 2.21   & 14.19 & 12.83  \\
% \bottomrule
% \end{tabular}}
% \label{t2}
% \end{table}
To explore the performance improvements brought by real data to the diarization module, we provide a brief comparison of the performance of diarization models trained with different datasets in Table \ref{tab:diaresult1}. As shown in the table, adding real training data in NOTSOFAR~\cite{chime8Meetingsrecordingsdataset} leads to a substantial improvement (DER from 21.51$\%$ to 16.52$\%$). 
\begin{table}[t]
	\renewcommand\arraystretch{1.25}
	\newcolumntype{L}[1]{>{\raggedright\arraybackslash}p{#1}}
	\newcolumntype{C}[1]{>{\centering\arraybackslash}p{#1}}
	\newcolumntype{R}[1]{>{\raggedleft\arraybackslash}p{#1}}
	\centering
	\caption{DER ($\%$) comparisons of different NSD training data sets on Dev-set-1 multi-channel track (without `rockfall$\_$1').}
	\vspace{-0.3cm}
	\label{tab:diaresult1}\medskip
	\resizebox{8 cm}{!}{\begin{tabular}{l|c}
			%\hline
			\toprule[1 pt]
			%\hline
			Training Data Sets&DER ($\%$)\\
			\midrule
			%\hline
			LibriSpeech Simulated Data + NOTSOFAR Simulated Data& 21.51\\
			+ Train-set-1/2 MC (split into single channel) &16.52\\
			\bottomrule[1 pt]
	\end{tabular}}
	%\vspace{-0.3 cm}
\end{table}
\subsubsection{Separation}
Table \ref{tab:sepresults} presents the results of different speech separation methods with a fixed back-end recognition model. From this table, we can observe that adding a classification network for overlapping segment detection brings some improvement to the speech separation results (from 26.68$\%$ to 25.14$\%$). Additionally, JDS shows a noticeable improvement compared to the CSS method (from 25.14$\%$ to 20.62$\%$), primarily due to its ability to utilize more accurate speaker time boundaries. Furthermore, we can improve performance further (from 20.62$\%$ to 20.29$\%$) by using speaker boundaries trained on more data (referred to as `Dia$\_$Mask' in the table) instead of only relying on the speaker time boundaries from JDS (the `Time-Mask' in Fig \ref{fig:4}). During the decoding process, we can also select matching CSD RTTM to segment the speech separation results. Since CSD results may have lower confusion errors, this can positively impact recognition results (from 20.29$\%$ to 19.95$\%$). Finally, extending the window length of JDS from 3 seconds to 8 seconds leads to further improvements in speech separation performance.
\begin{table}[t]
	\renewcommand\arraystretch{1.25}
	\newcolumntype{L}[1]{>{\raggedright\arraybackslash}p{#1}}
	\newcolumntype{C}[1]{>{\centering\arraybackslash}p{#1}}
	\newcolumntype{R}[1]{>{\raggedleft\arraybackslash}p{#1}}
	\centering
	\caption{TcpWER ($\%$) comparisons of different separation method on Train-set-1 multi-channel track (plaza$\_$0). The ASR model is based on original Whisper-large-v3. The separation training dataset is NOTSOFAR simulated data \cite{chime8simulate}.}
	\vspace{-0.3cm}
	\label{tab:sepresults}\medskip
	\resizebox{7.5 cm}{!}{\begin{tabular}{l|c}
			%\hline
			\toprule[1 pt]
			%\hline
			Separation Methods&TcpWER ($\%$)\\
			\midrule
			%\hline
			CSS (3-Sec) + MVDR& 26.68\\
			CSS (3-Sec) + Overlap Detection (3-Sec) + MVDR&25.14\\
			JDS (3-Sec) + MVDR&20.62\\
			JDS (3-Sec) + Dia$\_$Mask + MVDR &20.29\\
			JDS (3-Sec) + Dia$\_$Mask + CSD$\_$RTTM + MVDR &19.95\\
			JDS (8-Sec) + CSD$\_$RTTM + MVDR &18.57\\
			JDS (8-Sec) + Dia$\_$Mask + CSD$\_$RTTM + MVDR &17.47\\
			%Whisper large v3 + RPE + MOE 
			\bottomrule[1 pt]
	\end{tabular}}
	%\vspace{-0.3 cm}
\end{table}

\subsubsection{Speech Recognition}
Table \ref{tab:asrresults2} presents the performance improvements achieved through various modifications to the speech recognition model architecture. The term `Long Prompt' refers to using the previous decoding history as a decoder prompt, which follows the methods used in Whisper. The results indicate that both the MOE and RPE methods effectively enhance speech recognition performance. Additionally, incorporating WavLM features further improves the performance of the speech recognition model.
\begin{table}[t]
	\renewcommand\arraystretch{1.25}
	\newcolumntype{L}[1]{>{\raggedright\arraybackslash}p{#1}}
	\newcolumntype{C}[1]{>{\centering\arraybackslash}p{#1}}
	\newcolumntype{R}[1]{>{\raggedleft\arraybackslash}p{#1}}
	\centering
	\caption{TcpWER ($\%$) comparisons of different ASR models on Dev-set-1 (Oracle GSS using RTTM label). The training sets is `Train-set-1/2 MC GSS`.}
	\vspace{-0.3cm}
	\label{tab:asrresults2}\medskip
	\resizebox{6.5 cm}{!}{\begin{tabular}{l|c}
			%\hline
			\toprule[1 pt]
			%\hline
			Models&TcpWER ($\%$)\\
			\midrule
			%\hline
			Whisper large v3& 8.46\\
			Whisper large v3 + RPE&8.42\\
			Whisper large v3 + MOE&8.34\\
                + Timestamp (as showed in Table~\ref{tab:data})&8.25\\
                + RPE + Long Prompt + Noise KLD&7.61\\
                + WavLM&7.50\\
                %Whisper large v3 + RPE + MOE 
			\bottomrule[1 pt]
	\end{tabular}}
	%\vspace{-0.3 cm}
\end{table}
Table \ref{tab:asrresults1} shows the results of the backend ASR with different training datasets. `All-set MC GSS/NN' means the sum of `Train-set-1/2 MC NN/GSS' and `Dev-set-1 MC NN/GSS' in Table \ref{tab:data}. It demonstrates that real training data in NOTSOFAR~\cite{chime8Meetingsrecordingsdataset}, processed through oracle GSS, can significantly improve the performance of the backend ASR.
\begin{table}[t]
	\renewcommand\arraystretch{1.25}
	\newcolumntype{L}[1]{>{\raggedright\arraybackslash}p{#1}}
	\newcolumntype{C}[1]{>{\centering\arraybackslash}p{#1}}
	\newcolumntype{R}[1]{>{\raggedleft\arraybackslash}p{#1}}
	\centering
	\caption{TcpWER ($\%$) comparisons of Whisper large v3 models with different training data sets on Dev-set-2 (Oracle GSS using RTTM label). Note that we removed the anomalous session `MTG-30522'.}
	\vspace{-0.3cm}
	\label{tab:asrresults1}\medskip
	\resizebox{8 cm}{!}{\begin{tabular}{l|c}
			%\hline
			\toprule[1 pt]
			%\hline
			Training Data Set&TcpWER ($\%$)\\
			\midrule
			%\hline
			Original Datasets& 16.57\\
			Original Datasets + Train-set-1/2 MC GSS&12.07\\
			All-set MC GSS/NN + LibriSpeech Simulated Data&9.87\\
			\bottomrule[1 pt]
	\end{tabular}}
	%\vspace{-0.3 cm}
\end{table}
%\subssection{Single-channel System}
% In our final submission, sub-sys1 (no-diar,ASR-V1), main-sys1 (DER-P-diar,ASR-V1), and main-sys2 (WER-P-diar,ASR-V1) are consistent with the results as shown in Table~\ref{tab:da_diar}. 
% Given that the rules allow us to re-arrange the training set and development set, we move 80\% utterances of the development set into the official training set, and perform the same data augmentation process as stated in Section~\ref{Data_Augmentation}. In this way, we only retrain our end-to-end ASR models (ASR-V2), and submit corresponding results as sub-sys2 (no-diar,ASR-V2) and main-sys3 (WER-P-diar,ASR-V2).
\section{Conclusion}
The NOTSOFAR-1 challenge explored a meaningful scenario, namely real-world far-field multi-speaker meeting environments. This includes many challenges that speech signal processing systems need to deal with in practical applications, including speaker movement, high speech overlap rates, rapid changes in speakers, various noise and reverberation, and a variable number of speakers. In order to deal with these challenges, we proposed some methods from the front-end signal processing and back-end speech recognition, mainly including the use of data-driven NSD models to predict speaker time boundaries, combining traditional spatial information-based GSS and data-driven JDS models for speech separation, as well as the construction of speech recognition training data and the modification of speech recognition model architecture. 
In this challenge, we found that the NSD method requires effectively matched training data to improve performance, including both real and simulated datasets. Additionally, multi-stage optimization of the diarization priors of NSD proved to be an important factor for diarization performance. Furthermore, incorporating time boundary information from diarization can help the speech separation model achieve better separation results with more accurate time boundaries, thereby effectively improving speech recognition performance. Finally, fine-tuning with matched datasets and improving model architecture are still crucial methods for enhancing speech recognition performance. In the NOTSOFAR-1 challenge, our system achieved the tcpWERs of 22.2$\%$ and 10.8$\%$ in the single-channel and multi-channel tracks of the evaluation set, respectively, winning first place in both tracks.
\section{Acknowledgements}
This work was supported by the National Natural Science Foundation of China under Grants No. 62171427.
\bibliographystyle{IEEEtran}
\bibliography{mybib}

\begin{thebibliography}{10}
\providecommand{\url}[1]{#1}
\csname url@samestyle\endcsname
\providecommand{\newblock}{\relax}
\providecommand{\bibinfo}[2]{#2}
\providecommand{\BIBentrySTDinterwordspacing}{\spaceskip=0pt\relax}
\providecommand{\BIBentryALTinterwordstretchfactor}{4}
\providecommand{\BIBentryALTinterwordspacing}{\spaceskip=\fontdimen2\font plus
\BIBentryALTinterwordstretchfactor\fontdimen3\font minus
  \fontdimen4\font\relax}
\providecommand{\BIBforeignlanguage}[2]{{%
\expandafter\ifx\csname l@#1\endcsname\relax
\typeout{** WARNING: IEEEtran.bst: No hyphenation pattern has been}%
\typeout{** loaded for the language `#1'. Using the pattern for}%
\typeout{** the default language instead.}%
\else
\language=\csname l@#1\endcsname
\fi
#2}}
\providecommand{\BIBdecl}{\relax}
\BIBdecl

\bibitem{vinnikov24_interspeech}
A.~Vinnikov, A.~Ivry, A.~Hurvitz, I.~Abramovski, S.~Koubi, I.~Gurvich, S.~Peer,
  X.~Xiao, B.~M. Elizalde, N.~Kanda, X.~Wang, S.~Shaer, S.~Yagev, Y.~Asher,
  S.~Sivasankaran, Y.~Gong, M.~Tang, H.~Wang, and E.~Krupka, ``Notsofar-1
  challenge: New datasets, baseline, and tasks for distant meeting
  transcription,'' in \emph{Interspeech 2024}, 2024, pp. 5003--5007.

\bibitem{chime8basline}
``{CHiME-8 Baseline System},''
  \emph{\url{https://www.chimechallenge.org/current/task2/baseline}}, 2024.

\bibitem{chime8simulate}
``{CHiME-8 Simulated Training Dataset},''
  \emph{\url{https://www.chimechallenge.org/current/task2/datasimulated-training-dataset}},
  2024.

\bibitem{higuchi2017online}
T.~Higuchi, N.~Ito, S.~Araki, T.~Yoshioka, M.~Delcroix, and T.~Nakatani,
  ``{Online MVDR beamformer based on complex Gaussian mixture model with
  spatial prior for noise robust ASR},'' \emph{IEEE/ACM Transactions on Audio,
  Speech, and Language Processing}, vol.~25, no.~4, pp. 780--793, 2017.

\bibitem{nagrani2020voxceleb}
A.~Nagrani, J.~S. Chung, W.~Xie, and A.~Zisserman, ``Voxceleb: Large-scale
  speaker verification in the wild,'' \emph{Computer Speech \& Language},
  vol.~60, p. 101027, 2020.

\bibitem{MAMSE}
M.-K. He, J.~Du, Q.-F. Liu, and C.-H. Lee, ``{ANSD-MA-MSE: Adaptive Neural
  Speaker Diarization Using Memory-Aware Multi-Speaker Embedding},''
  \emph{IEEE/ACM Transactions on Audio, Speech, and Language Processing},
  vol.~31, pp. 1561--1573, 2023.

\bibitem{yang2024neural}
G.~Yang, M.~He, S.~Niu, R.~Wang, Y.~Yue, S.~Qian, S.~Wu, J.~Du, and C.-H. Lee,
  ``Neural speaker diarization using memory-aware multi-speaker embedding with
  sequence-to-sequence architecture,'' in \emph{ICASSP 2024-2024 IEEE
  International Conference on Acoustics, Speech and Signal Processing
  (ICASSP)}.\hskip 1em plus 0.5em minus 0.4em\relax IEEE, 2024, pp.
  11\,626--11\,630.

\bibitem{Wang2023TheUS}
\BIBentryALTinterwordspacing
R.~Wang, M.~He, J.~Du, H.~Zhou, S.~Niu, H.~Chen, Y.~Yue, G.~Yang, S.~Wu,
  L.~Sun, Y.~Tu, H.~Tang, S.~Qian, T.~Gao, M.~Wang, G.~Wan, J.~Pan, J.~Gao, and
  C.-H. Lee, ``The ustc-nercslip systems for the chime-7 dasr challenge,''
  \emph{ArXiv}, vol. abs/2308.14638, 2023. [Online]. Available:
  \url{https://api.semanticscholar.org/CorpusID:261244449}
\BIBentrySTDinterwordspacing

\bibitem{Radford2022RobustSR}
\BIBentryALTinterwordspacing
A.~Radford, J.~W. Kim, T.~Xu, G.~Brockman, C.~McLeavey, and I.~Sutskever,
  ``Robust speech recognition via large-scale weak supervision,'' \emph{ArXiv},
  vol. abs/2212.04356, 2022. [Online]. Available:
  \url{https://api.semanticscholar.org/CorpusID:252923993}
\BIBentrySTDinterwordspacing

\bibitem{Chen2021WavLMLS}
\BIBentryALTinterwordspacing
S.~Chen, C.~Wang, Z.~Chen, Y.~Wu, S.~Liu, Z.~Chen, J.~Li, N.~Kanda,
  T.~Yoshioka, X.~Xiao, J.~Wu, L.~Zhou, S.~Ren, Y.~Qian, Y.~Qian, M.~Zeng, and
  F.~Wei, ``Wavlm: Large-scale self-supervised pre-training for full stack
  speech processing,'' \emph{IEEE Journal of Selected Topics in Signal
  Processing}, vol.~16, pp. 1505--1518, 2021. [Online]. Available:
  \url{https://api.semanticscholar.org/CorpusID:239885872}
\BIBentrySTDinterwordspacing

\bibitem{Jiang2022NextformerAC}
\BIBentryALTinterwordspacing
Y.~Jiang, J.~Yu, W.~Yang, B.~Zhang, and Y.~Wang, ``Nextformer: A convnext
  augmented conformer for end-to-end speech recognition,'' 2022. [Online].
  Available: \url{https://api.semanticscholar.org/CorpusID:250113612}
\BIBentrySTDinterwordspacing

\bibitem{Zhou2019ImprovingGO}
\BIBentryALTinterwordspacing
P.~Zhou, R.~Fan, W.~Chen, and J.~Jia, ``Improving generalization of transformer
  for speech recognition with parallel schedule sampling and relative
  positional embedding,'' \emph{ArXiv}, vol. abs/1911.00203, 2019. [Online].
  Available: \url{https://api.semanticscholar.org/CorpusID:207870654}
\BIBentrySTDinterwordspacing

\bibitem{Dai2021CoAtNetMC}
\BIBentryALTinterwordspacing
Z.~Dai, H.~Liu, Q.~V. Le, and M.~Tan, ``Coatnet: Marrying convolution and
  attention for all data sizes,'' \emph{ArXiv}, vol. abs/2106.04803, 2021.
  [Online]. Available: \url{https://api.semanticscholar.org/CorpusID:235376986}
\BIBentrySTDinterwordspacing

\bibitem{Sun2023RetentiveNA}
\BIBentryALTinterwordspacing
Y.~Sun, L.~Dong, S.~Huang, S.~Ma, Y.~Xia, J.~Xue, J.~Wang, and F.~Wei,
  ``Retentive network: A successor to transformer for large language models,''
  \emph{ArXiv}, vol. abs/2307.08621, 2023. [Online]. Available:
  \url{https://api.semanticscholar.org/CorpusID:259937453}
\BIBentrySTDinterwordspacing

\bibitem{Peng2023RWKVRR}
\BIBentryALTinterwordspacing
B.~Peng, E.~Alcaide, Q.~G. Anthony, A.~Albalak, S.~Arcadinho, S.~Biderman,
  H.~Cao, X.~Cheng, M.~Chung, M.~Grella, G.~Kranthikiran, X.~He, H.~Hou,
  P.~Kazienko, J.~Kocoń, J.~Kong, B.~Koptyra, H.~Lau, K.~S.~I. Mantri, F.~Mom,
  A.~Saito, X.~Tang, B.~Wang, J.~S. Wind, S.~Wozniak, R.~Zhang, Z.~Zhang,
  Q.~Zhao, P.~Zhou, J.~Zhu, and R.~Zhu, ``Rwkv: Reinventing rnns for the
  transformer era,'' in \emph{Conference on Empirical Methods in Natural
  Language Processing}, 2023. [Online]. Available:
  \url{https://api.semanticscholar.org/CorpusID:258832459}
\BIBentrySTDinterwordspacing

\bibitem{Gulati2020ConformerCT}
\BIBentryALTinterwordspacing
A.~Gulati, J.~Qin, C.-C. Chiu, N.~Parmar, Y.~Zhang, J.~Yu, W.~Han, S.~Wang,
  Z.~Zhang, Y.~Wu, and R.~Pang, ``Conformer: Convolution-augmented transformer
  for speech recognition,'' \emph{ArXiv}, vol. abs/2005.08100, 2020. [Online].
  Available: \url{https://api.semanticscholar.org/CorpusID:218674528}
\BIBentrySTDinterwordspacing

\bibitem{You2021SpeechMoEST}
\BIBentryALTinterwordspacing
Z.~You, S.~Feng, D.~Su, and D.~Yu, ``Speechmoe: Scaling to large acoustic
  models with dynamic routing mixture of experts,'' in \emph{Interspeech},
  2021. [Online]. Available:
  \url{https://api.semanticscholar.org/CorpusID:234094030}
\BIBentrySTDinterwordspacing

\bibitem{chime8Meetingsrecordingsdataset}
``{CHiME-8 Meetings Recordings Dataset},''
  \emph{\url{https://www.chimechallenge.org/current/task2/datasimulated-training-dataset}},
  2024.

\bibitem{musan2015}
D.~Snyder, G.~Chen, and D.~Povey, ``{MUSAN}: {A} {M}usic, {S}peech, and {N}oise
  {C}orpus,'' 2015, arXiv:1510.08484v1.

\bibitem{Ji2023ResearchOA}
\BIBentryALTinterwordspacing
W.~Ji, S.~Zan, G.~Zhou, and X.~Wang, ``Research on an improved conformer
  end-to-end speech recognition model with r-drop structure,'' \emph{ArXiv},
  vol. abs/2306.08329, 2023. [Online]. Available:
  \url{https://api.semanticscholar.org/CorpusID:259165369}
\BIBentrySTDinterwordspacing

\bibitem{wang2024incorporatingspatialcuesmodular}
\BIBentryALTinterwordspacing
R.~Wang, S.~Niu, G.~Yang, J.~Du, S.~Qian, T.~Gao, and J.~Pan, ``Incorporating
  spatial cues in modular speaker diarization for multi-channel multi-party
  meetings,'' 2024. [Online]. Available: \url{https://arxiv.org/abs/2409.16803}
\BIBentrySTDinterwordspacing

\end{thebibliography}

\end{document}